\documentclass[reprint,superscriptaddress, amsmath,amssymb,pra, aps,floatfix,longbibliography]{revtex4-1}
\usepackage[utf8]{inputenc}

\usepackage{hyperref}% add hypertext capabilities
\usepackage{dcolumn}% Align table columns on decimal point
\usepackage{bm}% bold math
\usepackage{hyperref}% add hypertext capabilities
\usepackage[dvipsnames]{xcolor}
\usepackage{graphicx}
\graphicspath{ {./images/} }
\usepackage[capitalise]{cleveref}
\usepackage{textcomp}
\usepackage{pdfcomment}
\usepackage{bold-extra}
\usepackage{upgreek}
\newcommand{\ks}[1]{\textcolor{blue}{#1}}

\DeclareUnicodeCharacter{2009}{\,}
\hypersetup{pdftoolbar=true,        % show Acrobat’s toolbar?
 pdfmenubar=true,        % show Acrobat’s menu?
 pdffitwindow=false,     % window fit to page when opened
 pdfstartview={fitH},    % fits the width of the page to the window
 pdftitle={},    % title
 pdfauthor={Gregory Moille},     % author
 pdfsubject={DWsynthesis},   % subject of the document
 pdfcreator={Gregory Moille},   % creator of the document
 pdfproducer={LaTeX}, % producer of the document
 pdfkeywords={Version 1.0},
 colorlinks=true,       % false: boxed links; true: colored links
 linkcolor=blue,          % color of internal links (change box color with
 citecolor=blue,        % color of links to bibliography
 filecolor=blue,      % color of file links
 urlcolor=blue}

\makeatletter
\renewcommand{\@biblabel}[1]{#1. }
\renewcommand{\@dotsep}{500}
\renewcommand{\@pnumwidth}{0em}
\renewcommand{\l@figure}[2]{% #1 is e.g., Figure 1 + caption, #2 is pg.
\@dottedtocline{1}{1.5em}{2em}{Figure #1}{}\vspace{15pt}}

\usepackage[normalem]{ulem}

\begin{document}

\title{Full spectral response of grating-induced loss in photonic crystal microrings}

\author{Daniel Pimbi}   
\thanks{These two authors contributed equally}
\affiliation{Microsystems and Nanotechnology Division, Physical Measurement Laboratory, National Institute of Standards and Technology, Gaithersburg, Maryland 20899, USA}
\affiliation{Joint Quantum Institute, NIST/University of Maryland, College Park, Maryland 20742, USA}

\author{Yi Sun}   
\thanks{These two authors contributed equally}
\affiliation{Microsystems and Nanotechnology Division, Physical Measurement Laboratory, National Institute of Standards and Technology, Gaithersburg, Maryland 20899, USA}
\affiliation{Joint Quantum Institute, NIST/University of Maryland, College Park, Maryland 20742, USA}

\author{Roy Zektzer}   % unless can't finish
\affiliation{Microsystems and Nanotechnology Division, Physical Measurement Laboratory, National Institute of Standards and Technology, Gaithersburg, Maryland 20899, USA}
\affiliation{Joint Quantum Institute, NIST/University of Maryland, College Park, Maryland 20742, USA}

\author{Xiyuan Lu}\email{xnl9@umd.edu}
\affiliation{Microsystems and Nanotechnology Division, Physical Measurement Laboratory, National Institute of Standards and Technology, Gaithersburg, Maryland 20899, USA}
\affiliation{Joint Quantum Institute, NIST/University of Maryland, College Park, Maryland 20742, USA}

\author{Kartik Srinivasan} \email{kartik.srinivasan@nist.gov}
\affiliation{Microsystems and Nanotechnology Division, Physical Measurement Laboratory, National Institute of Standards and Technology, Gaithersburg, Maryland 20899, USA}
\affiliation{Joint Quantum Institute, NIST/University of Maryland, College Park, Maryland 20742, USA}

\date{\today}% It is always \today, today,

\begin{abstract}
      Photonic crystal microrings (PhCRs) have emerged as powerful and versatile platforms for integrated nonlinear photonics, offering precise control over frequency and phase matching while maintaining high optical quality factors. Through grating-mediated mode coupling, PhCRs enable advanced dispersion engineering, which is critical for wideband nonlinear processes such as optical parametric oscillation, Kerr frequency comb generation, and dual-pump spontaneous and Bragg scattering four-wave mixing. Beyond dispersion control, PhCRs also facilitate the manipulation of orbital angular momentum (OAM) emission, a key functionality for encoding high-dimensional quantum states in emerging quantum photonic platforms. Despite these advances, the broadband spectral behavior of grating-induced losses in PhCRs remains largely unexplored, with most studies focusing on grating periods near the modal wavelength or its half. Such losses can significantly impact broadband nonlinear processes, where excess loss at unintended wavelengths can degrade device performance. In this work, we experimentally characterize grating-induced losses in PhCRs and reveal their full spectral response as a function of the ratio between modal wavelength and grating period. We identify distinct loss channels arising from either radiation or mode conversion, including a broad excess-loss region attributed to vertical out-coupling into OAM-carrying states. These observations are supported by three-dimensional finite-difference time-domain simulations and further analyzed through OAM radiation angle and phase-mismatch analysis. The resulting broadband loss spectrum highlights critical design trade-offs and provides practical guidelines for optimizing PhCR-based devices for nonlinear photonic applications involving widely separated frequencies.
\end{abstract}
% \pacs{78.67.Hc, 42.70.Qs, 42.60.Da} 

\maketitle
\section{Introduction}
\noindent The incorporation of gratings into microring resonators ~\cite{Lee_OL_2012, zhang_resonance-splitting_2008, lu_selective_2014} introduces new methods for controlling the spectral and spatial properties of high-quality-factor (Q) whispering-gallery modes~\cite{Lu_Nanophotonics_2023}. These grating-inscribed devices, which are now commonly referred to as photonic crystal microrings (PhCRs), have been widely explored for single-wavelength-band applications such as optical filters~\cite{li_using_2019, wu_side_mode_suppressed_2019}, sensing \cite{leonardis_performance_2014, lo_phc_biosensing_2017, ma_integrated_label_free_2016}, and lasers~\cite{arbabi_realization_2011, arbabi_grating_single_mode_laser_2015}. More recently, PhCRs have gained significant attention in nonlinear photonics, where their ability to selectively control individual mode frequencies enables precise frequency and phase matching for processes such as optical parametric oscillation (OPO)~\cite{Black_Optica_2022,lu_optics_letters_2022, Stone_NatPhoton_2024, li_broadband_and_accurate_2025}, frequency comb generation ~\cite{Lucas_NatPhoton_2023, liu_implementing_2025}, including its vertical vortex emission~\cite{liu_integrated_vortex_2024, chen_integrated_vortex_2024} and spontaneous soliton formation~\cite{Yu_NatPhoton_2021, yu_continuum_bright_pulse_2022}, self-injection-locked lasing~\cite{Ulanov_NatPhoton_2024, lu_band_flipping_2024}, dual-pump spontaneous and Bragg scattering four-wave mixing~\cite{ulanov_quadrature_squeezing_2025}, and stimulated Brillouin scattering~\cite{Wang_PRX_2024}. In these applications, illustrated in Fig. 1(a), the inscribed gratings serve as key enablers, providing orbital angular momentum (OAM) emission~\cite{cai_integrated_oam_2012, willner_OAM_2021,Wang_PRL_2022,Lu_NatCommun_2023}, single or multiple-mode splitting~\cite{Lu_PhotonRes_2020}, spectral selectivity~\cite{Lu_NatPhoton_2022}, and engineered dispersion~\cite{Lu_PhotonRes_2023,Moille_CommunPhys_2023,Lucas_NatPhoton_2023} that are challenging to achieve with standard microring geometries alone.

\begin{figure*}[ht!]
\centering\includegraphics[width=0.98\linewidth]{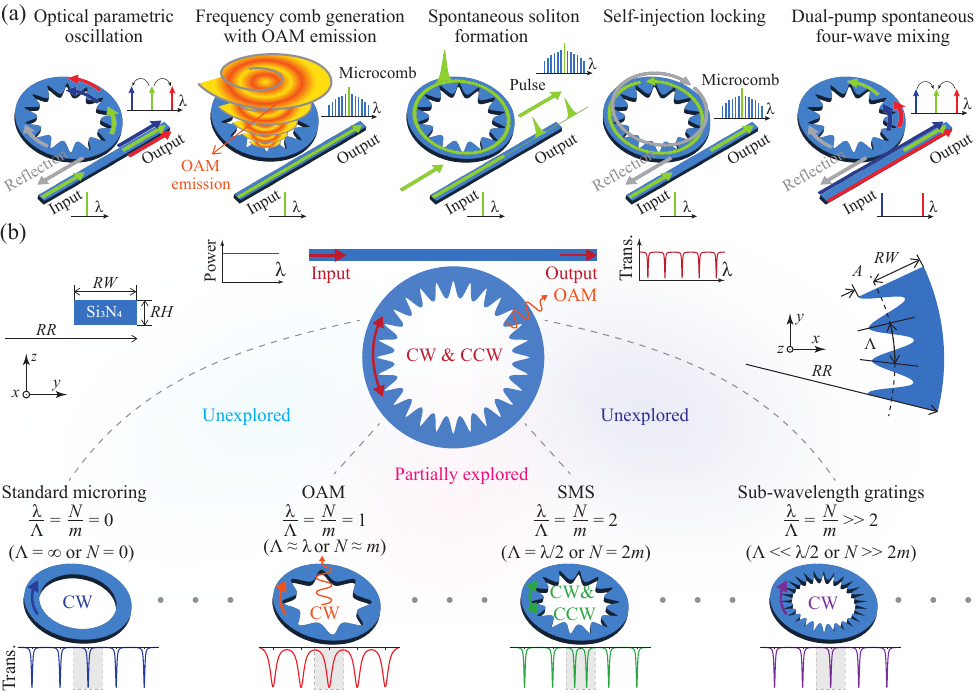}
\caption{Applications of photonic crystal microring resonators (PhCRs) and an approach for characterizing their broadband spectrum of grating-induced loss without requiring multiple laser sources. (a) Selected examples of nonlinear photonic applications enabled by PhCRs, including optical parametric oscillation (OPO)~\cite{Black_Optica_2022, Stone_NatPhoton_2024, Lu_PhotonRes_2023}, Kerr frequency comb generation~~\cite{Moille_CommunPhys_2023, liu_implementing_2025} with orbital angular momentum (OAM) emission~\cite{liu_integrated_vortex_2024, chen_integrated_vortex_2024}, spontaneous soliton formation~\cite{Yu_NatPhoton_2021, yu_continuum_bright_pulse_2022}, self-injection locking~\cite{Ulanov_NatPhoton_2024, lu_band_flipping_2024}, and dual-pump spontaneous four-wave mixing (DP-SFWM)~\cite{ulanov_quadrature_squeezing_2025}. In each application, the incorporation of a grating into the microring introduces an additional degree of freedom for controlling frequency and phase matching conditions, thereby facilitating these nonlinear processes. (b) Method for studying grating-induced loss in PhCRs by varying the grating period ($\Lambda$), that is, the number of inscribed grating periods ($N$), in a microring of fixed circumference while keeping all other parameters constant. We consider a PhCR with gratings inscribed along the inner boundary, as shown at the top. The left and right inset schematics respectively depict a cross-sectional view of the microring cross-section and a zoomed-in top view highlighting key geometric parameters: ring radius ($RR$), ring width ($RW$), thickness ($RH$), grating period ($\Lambda$), and modulation amplitude ($A$). The bottom panel illustrates key values of the grating period $\Lambda$ (or equivalently $N$) relative to the modal wavelength ($\lambda$) and the azimuthal mode number ($m$), highlighting well-understood conditions as well as intermediate regions where the loss mechanisms have thus far been partially or entirely unexplored. By varying $\Lambda$ across multiple devices and studying the resulting impact on loss at a fixed $\lambda$, we can extract the loss spectrum at fixed $\Lambda$ and varying $\lambda$.
} 
\label{Fig1}
\end{figure*}

While these capabilities have expanded the role of PhCRs in nonlinear photonics, the broader spectral implications of grating-induced losses remain incompletely understood. This limited understanding is primarily due to the need for broadband tunable laser sources to probe the full spectral response across nearly an octave or more of spectral bandwidth. Although multiple narrow-band sources can be used to span wider spectral ranges, such approaches are often experimentally cumbersome, power-limited, and lack sufficient spectral resolution. Moreover, grating-induced losses are frequently subtle and may be masked by other wavelength-dependent mechanisms such as material absorption~\cite{luke_si3n4_2015, nitkowski_cavity_enhanced_2008}, surface scattering~\cite{sorayaie_focus_on_surface_scattering_2025, lee_influence_of_surface_2019}, or coupling inefficiencies~\cite{moille_broadband_CMT_2019, cui_distinguishing_2024}. As a result, most previous studies have focused on a few well-established configurations, as shown in Fig. 1(b). Here, the length of the grating period $\Lambda$ in comparison to the modal wavelength $\lambda = \lambda_{0}/n_\text{eff}$, where $\lambda_{0}$ is the free-space wavelength and $n_\text{eff}$ is the effective modal refractive index, is important in demarcating different operating regions. The standard microring without gratings ($\Lambda = \infty$ or the number of periods $N = 0$) exhibits no grating-induced loss, but instead an intrinsic loss mainly due to material and geometric parameters~\cite{ji_methods_2021}. When $\Lambda$ is near the modal wavelength ($\Lambda \approx \lambda$), vertical emission into OAM modes introduces noticeable excess loss~\cite{Lu_NatCommun_2023}. This condition is equivalent to $N$ being nearly equal to the microring azimuthal mode number ($N \approx m$). When the period is half the modal wavelength ($\Lambda = \lambda/2$ or $N = 2m$), it satisfies the coherent backscattering condition~\cite{Stone_NatPhoton_2024}, also referred to as the Bragg condition~\cite{Yariv}, enabling coupling between clockwise (CW) and counterclockwise (CCW) propagating modes and producing single-mode frequency splitting with minimal additional loss~\cite{lu_selective_2014,Lu_PhotonRes_2020}. In the deep subwavelength regime ($\Lambda \ll \lambda/2$ or $N \gg 2m$), waveguide gratings are known to suppress both diffraction and reflection effects~\cite{flueckiger_subwavelength_grating_2016}, suggesting that PhCRs operating in this regime should have losses comparable to those of standard microrings, though this is yet to be experimentally validated. Finally, in the intermediate regions ($\infty>\Lambda>\lambda$, $\lambda>\Lambda>\lambda/2$, and $\lambda/2>\Lambda$) grating-induced loss remains unexplored or only partially characterized~\cite{Lu_NatCommun_2023}. A clear understanding of the behavior in-between the well-characterized regions is important for nonlinear processes involving widely separated frequencies. In particular, the corresponding wavelengths can significantly differ relative to $\Lambda$, so that a single device may need to function across multiple regions. Unintended losses at relevant wavelengths can degrade performance by increasing threshold powers and reducing conversion efficiency.

In this work, we investigate the full spectral response of grating-induced loss in PhCRs and establish a quantitative relationship between the resonator quality factor Q, modal wavelength $\lambda$, and grating period $\Lambda$. We begin by experimentally characterizing grating-induced losses in PhCR devices with varying $\Lambda$, corresponding to $N$ ranging from 0 to 600, inscribed within a microring of fixed circumference. These measurements were performed across 37 devices at a fixed free-space wavelength ($\lambda_0 \approx 1550$~nm) and azimuthal mode number ($m = 165$). We identify distinct loss channels that depend on the ratio $\lambda/\Lambda$, appearing within a broad excess loss region attributed to vertical out-coupling into states carrying OAM. Our experimental findings are supported by three-dimensional finite-difference time-domain (3D FDTD) simulations, and the origins of the loss channels are further revealed through analysis of the OAM radiation angle and modal phase mismatch. Building on these results, we numerically study the reverse problem, that is, how a PhCR with a fixed $\Lambda$ operates across a broad wavelength range. This broadband response is then used to develop design guidelines for minimizing losses and enhancing PhCR performance in two different nonlinear integrated photonics applications.

\begin{figure*}[t!]
\centering\includegraphics[width=0.97\linewidth]{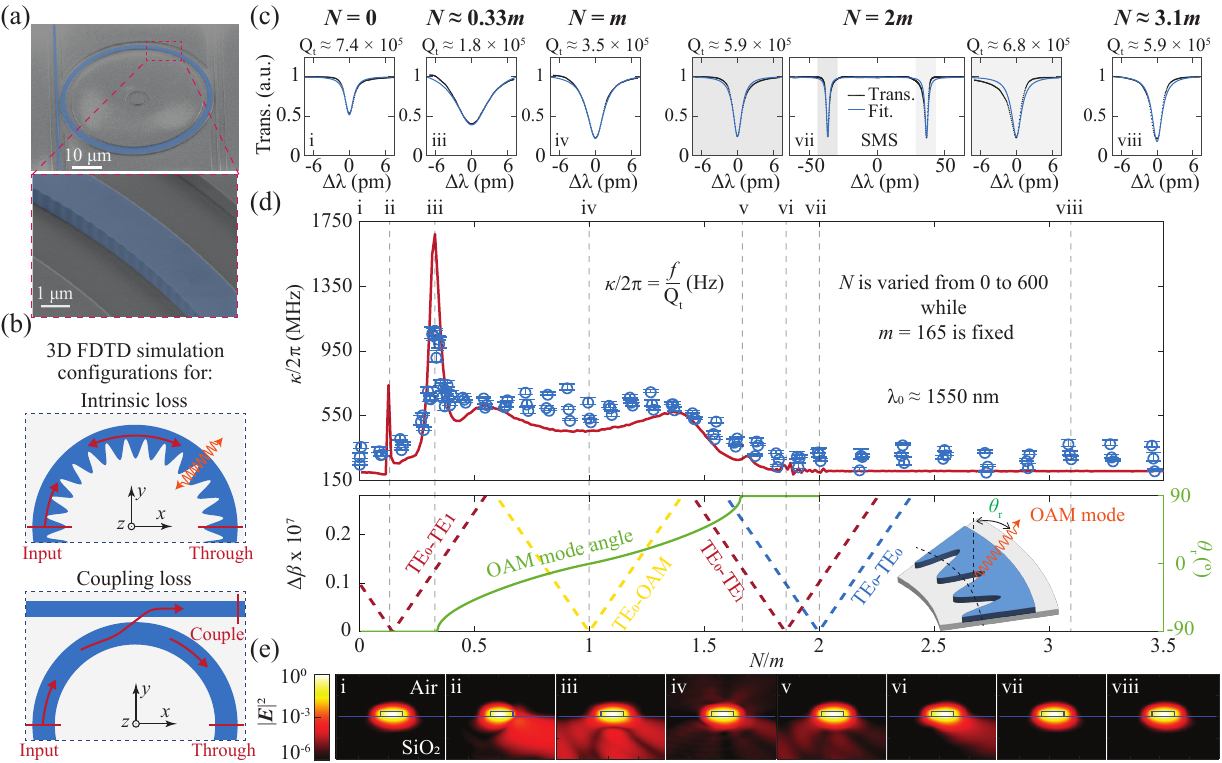}
\caption{Experimental characterization of grating-induced loss in PhCRs, supported by 3D FDTD simulations. (a) Scanning electron microscope (SEM) image of a fabricated PhCR device (top) and a magnified view of the inscribed grating (bottom), with false-color blue indicating the Si$_3$N$_4$ waveguide. (b) Top-view schematics of the 3D FDTD simulation setups. The top panel shows the configuration used to compute grating-induced loss using a half PhCR without a bus waveguide. The bottom panel shows the configuration used to calculate coupling loss in a standard microring. (c) Experimental transmission spectra (black) with corresponding nonlinear least-squares fit (blue) for $N = 0$, $0.33m$, $m$, $2m$, and $3.1m$, measured at a fixed telecom wavelength $\lambda_0 = 1550$~nm with azimuthal mode number $m = 165$. The extracted total quality factors ($Q_\text{t}$) from each fit are given above the corresponding resonances. At $N = 2m$, single-mode splitting (SMS) is observed, resulting from coherent backscattering between counterpropagating degenerate TE$_0$ modes. (d) Total radiation loss ($\kappa$) as a function of the grating-to-mode number ratio ($N/m$) for values of $N$ ranging from 0 to 600, across 37 devices. Experimental loss values (blue circles), extracted from the transmission spectra in (a), are overlaid with 3D FDTD simulation results (solid red trace). Error bars represent one standard deviation from the nonlinear least-squares fitting. The dashed vertical lines (i) through (viii) correspond to $N \approx 0$, $0.13m$, $0.33m$, $m$, $1.67m$, $1.87m$, $2m$, and $3.1m$, respectively. The bottom panel shows the calculated radiation angle ($\theta_{\text{r}}$) for OAM modes (solid green line), along with the relative phase mismatch ($\Delta\beta$) for different coupling scenarios: TE$_0$–TE$_1$ mode coupling (red dashed line), TE$_0$–OAM mode coupling (yellow dashed line), and TE$_0$–TE$_0$ mode coupling (blue dashed line). Identified loss channels include OAM radiation centered at region iv, surface-mode radiation at regions iii and v, TE$_0$–TE$_1$ mode forward/backward coupling at regions ii and vi, and negligible excess loss at regions i and viii, including TE$_0$–TE$_0$ backscattering at region vii. The inset plot shows a 3D view of the PhCR waveguide with inscribed grating, illustrating the radiation angle of OAM modes. (e) FDTD-simulated electric field intensity distributions for loss regions i–viii, taken at the PhCR cross-section, illustrating the origin of different observed radiation loss channels.
} 
\label{Fig2}
\end{figure*}

\begin{figure*}[t!]
\centering\includegraphics[width=0.98\linewidth]{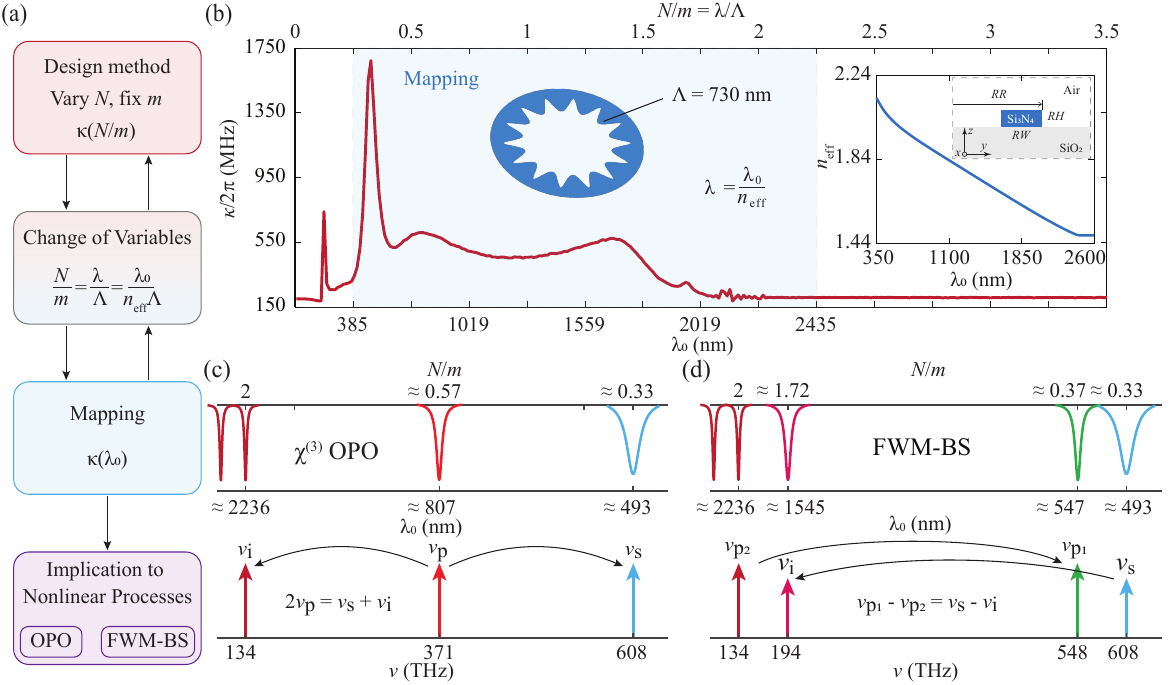}
\caption{A mathematical approach (a) for mapping the $N/m$-dependent grating-induced loss from Fig.~\ref{Fig2} to a full spectral response as a function of wavelength in (b), along with its implications to nonlinear photonic processes (c)-(d). (a) Mathematical mapping procedure. We first study PhCRs by varying the number of grating periods ($N$) while keeping the azimuthal mode number ($m$) fixed (see Fig.~\ref{Fig2}(c–d)). The variables are transformed using the relationships $\Lambda = 2\pi (RR - RW)/N$ and $\lambda = 2\pi (RR - RW)/m = \lambda_0 / (n_\text{eff} \Lambda)$, where $\lambda_0$ is the free-space wavelength and $n_\text{eff}$ is the effective refractive index. The loss as a function of the ratio $N/m = \lambda/\Lambda$ is then mapped to a function of $\lambda_0$. (b) Full spectral response of grating-induced loss for a single PhCR with a fixed grating period $\Lambda = 730$~nm. The top x-axis shows the transformation from the grating-to-mode number ratio ($N/m$) to the wavelength-to-grating-period ratio ($\lambda/\Lambda$). The bottom x-axis and shaded region represent the corresponding mapping to $\lambda_0$, using $\lambda = \lambda_0/n_\text{eff}$. (c–d) Examples of nonlinear processes where grating-induced radiation loss around $N/m \approx 0.33$ adversely impacts device performance. (c) OPO: A pump at 807~nm ($N/m = 0.57$) generates a signal at 493~nm ($N/m = 0.33$) and an idler at 2236~nm ($N/m = 2$). The modal splitting at $N/m = 2$ is used to satisfy frequency and phase matching, while the signal overlaps a radiation loss peak, reducing OPO efficiency. (d) FWM-BS: Two pumps at 547~nm ($N/m = 0.37$) and 2236~nm ($N/m = 2$) generate a signal at 608~nm ($N/m = 0.33$), which is subsequently downconverted to an idler at 1545~nm ($N/m = 1.72$). The signal wavelength aligns with a strong loss channel, potentially limiting conversion efficiency.
} 
\label{Fig3}
\end{figure*}

\section{Experimental and Numerical Analysis}

We fabricated our devices using a silicon nitride (Si$_3$N$_4$) layer on a silicon dioxide (SiO$_2$) lower cladding, with air cladding on the top and along the sides. Figure 2(a) shows scanning electron microscope (SEM) images of a fabricated PhCR device (top), along with a magnified view of the inscribed gratings (bottom). Each device has an outer radius of $RR = 25~\upmu$m, a ring width of $RW = 1500$~nm, and a thickness of $RH = 500$~nm, with a grating modulation amplitude of $A = 6$~nm. The number of grating periods is given by $N = 2\pi(RR - RW)/\Lambda$, and the azimuthal mode number is defined as $m = 2\pi(RR - RW)/\lambda$. Our experimental methodology for investigating losses in PhCR devices (illustrated in Fig.~1(b)) involves systematically varying the number of grating periods $N$ while keeping all other parameters constant, including the azimuthal mode number $m$. Because the PhCR operating regime depends on $\Lambda/\lambda$, this approach enables a controlled study of the broadband spectral response of PhCRs, analogous to a wavelength-swept analysis. This is accomplished without needing multiple broadband laser sources and varying $\lambda$, but instead using multiple devices and varying $\Lambda$. Figure~2(c) shows experimental transmission spectra (black lines) alongside nonlinear least-squares fits (blue lines) for selected values of $N$, all measured at a fixed azimuthal number of $m = 165$. These spectra all display singlet Lorentzian resonances except for a single mode splitting observed at $N = 2 m$, indicating strong coupling between clockwise (CW) and counterclockwise (CCW) propagating modes due to Bragg scattering. From each fit, we extracted the loaded quality factor $Q_\text{t}$ and used it to compute the total loss in units of frequency $f$ using the relation $\kappa/2\pi = f/Q_\text{t}$. The extracted losses from transmission measurements for all values of $N$, ranging from 0 to 600 across 37 devices, are plotted in Fig.~2(c) as blue circles. We observe no significant excess loss for large values of $N$ ($N\gtrsim2m$, regions vii and viii), relative to a microring without a grating ($N = 0$), including the coherent backscattering condition at $N = 2 m$ (region vii). In contrast, a broad region of excess loss centered around $N = m$ is evident, likely due to vertical out-coupling through states carrying OAM. Additionally, a pronounced loss peak appears at $N/m \approx 0.33$ (region iii), suggesting another radiation loss mechanism.

To validate our experimental observations, we performed numerical simulations of a half PhCR device using a 3D FDTD solver. This approach captures the essential physics of grating-induced loss mechanisms while significantly reducing computational time compared to full-ring modeling~\cite{ji_compact_spatial_mode_2022, bahadori_design_space_2018}. The simulations were divided into two configurations. The first configuration, shown at the top of Fig. 2(b), was used to calculate the intrinsic loss. It consists of a half PhCR structure with an input port that functions as both a mode source and a mode expansion monitor, and a through port that also serves as a mode expansion monitor. The simulation domain is enclosed with perfectly matched layers to suppress artificial reflections. A fundamental transverse electric mode (TE$_0$) was injected into the PhCR waveguide, and the transmitted and reflected powers of both the TE$_0$ and first-order transverse electric (TE$_1$) modes were recorded. The intrinsic loss was computed as the difference between the total input power and the sum of the transmitted and reflected modal powers, multiplied by two to account for the full PhCR device. These simulations were performed for a range of grating period numbers ($N = 0:1:600$) at a fixed free-space wavelength of 1550~nm. We set the grating modulation amplitude to $A = 20$~nm, dictated by the minimum Cartesian mesh resolution of 20~nm. The resulting loss values were subsequently rescaled to correspond to the experimental modulation amplitude of $A = 6$~nm. The second configuration, shown at the bottom of Fig. 2(b), was used to determine the coupling loss by simulating the microring-to-bus-waveguide region without any inscribed gratings. In this setup, a TE$_0$ mode was launched into the microring waveguide, and the TE$_0$ power coupled into the bus waveguide's coupling port was recorded. This coupling loss was then added to the intrinsic losses to obtain the total loss spectrum, shown as the solid red trace in Fig. 2(c). In addition to accurately reproducing peak iii and the broad excess loss observed between $N/m \approx 1.2$ and $N/m \approx 1.8$, the simulation also predicts the presence of peak ii, attributed to coupling between the TE$_0$ and TE$_1$ modes. It is important to note that these simulations were performed in a non-resonant ring waveguide configuration with a fine resolution in $N$. As such, the TE$_0$–TE$_1$ coupling peak is not observed in the experimental measurements—likely due to the non-degenerate nature of the mode frequencies or the relatively coarse discrete $N$ steps used in the fabricated devices. Additionally, smaller radiation loss peaks are observed at regions v and vi, which appear symmetrically positioned relative to peaks iii and ii with respect to the $N/m = 1$ axis.

\begin{figure}[t!]
\centering
\includegraphics[width=0.98\linewidth]{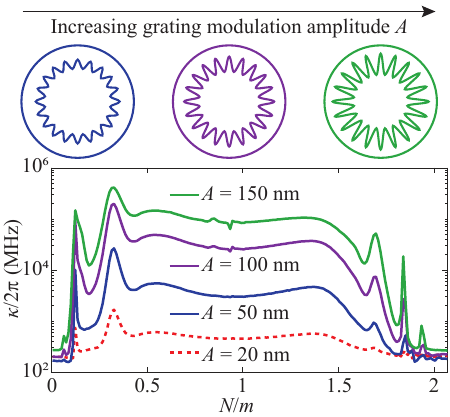}
\caption{Simulated grating-induced loss spectra as a function of the grating-to-mode number ratio ($N/m$) for grating modulation amplitudes of $A = 20$~nm (dashed red line), $A = 50$~nm (solid blue line), $A = 100$~nm (solid purple line), and $A = 150$~nm (solid green line). The dashed red line represents the same spectrum shown in Fig.~\ref{Fig2}(d). Here, all the spectra are plotted on a logarithmic scale. As the modulation amplitude increases, the overall loss magnitude increases accordingly; however, the spectral shape and the positions of the primary radiation loss channels remain largely unchanged.}
\label{fig5}
\end{figure}

We investigated the origins of the observed loss channels by analyzing the radiation angle ($\theta_{\text{r}}$) of OAM modes into the surrounding claddings (air and SiO$_2$), as well as the relative phase mismatch ($\Delta\beta$) between interacting waveguide modes. The radiation angle $\theta_{\text{r}}$ is obtained from Snell's law, $n_\text{e} \sin(\theta_{\text{e}}) = n_\text{r} \sin(\theta_{\text{r}})$~\cite{alonso_diffraction_less_2019}, where $\theta_{\text{e}} = (l/m)(\pi/2)$~\cite{Lu_NatCommun_2023} is the internal ejection angle of OAM modes with quantum number $l$ inside the silicon nitride PhCR waveguide. $n_\text{e}$ denotes the average effective index of the TE$_0$ and OAM modes, and $ n_\text{r}$ is the average refractive index of the top (air) and bottom (SiO$_2$) claddings. The phase mismatch is defined as $\Delta\beta = \beta_1 - \beta_2 - 2\pi/\Lambda$, where $\beta = 2\pi n_{\text{eff}}/\lambda$ is the propagation constant of each mode~\cite{Yariv}. Notice that both $\theta$ and $\Delta\beta$ depend on the effective refractive indices ($n_\text{eff}$); thus, we performed 2D eigenmode simulations of the ring waveguide cross-section. The bottom plot of Fig. 2(d) shows the calculated radiation angle $\theta_{\text{r}}$ for OAM modes (solid green line) and the relative phase mismatch $\Delta\beta$ for three mode combinations: TE$_0$–TE$_1$ (red dashed lines), TE$_0$–OAM (yellow dashed line), and TE$_0$–TE$_0$ (blue dashed line). From the radiation angle analysis, we identify that peak iv corresponds to the OAM mode centered at $N = m$, radiating at $\theta_{\text{r}} = 0^\circ$, and that peaks iii and v correspond to radiation along the glass/air interface at $\theta_{\text{r}} = \pm 90^\circ$ relative to the central OAM mode at $N=m$. The phase mismatch analysis reveals that peaks ii and vi are associated with TE$_0$–TE$_1$ mode coupling (forward and backward), while peak vii corresponds to TE$_0$–TE$_0$ backward Bragg reflection. Channels v and vi are notably suppressed under the air-clad condition compared to peaks iii and ii, but become more pronounced in simulations using oxide cladding, highlighting the sensitivity of these loss channels to the cladding environment (see appendix for details). Although these estimates are obtained using $n_\text{eff}$ data, they provide good predictions that closely match our experiments and full 3D FDTD simulation results. We further validated the origins of peaks ii through vi by analyzing electric field profiles obtained from cross-sections of the PhCR waveguide. As shown in Fig. 2(e), no significant radiation into the surrounding claddings was observed for channels vii and viii, as well as the reference case without an inscribed grating (region i). In contrast, channels ii through vi exhibited evident electric field leakage into the claddings, consistent with the excess radiation losses predicted in Fig. 2(d). These results confirm the proposed physical mechanisms responsible for the grating-induced radiation losses observed in both experiment and simulation.

Next, we studied numerically how a single PhCR device with a fixed grating period ($\Lambda = 730$~nm) operates across a broad spectrum, bounded at shorter wavelengths by the Si$_3$N$_4$ transparency limit ($\lambda_0 \lesssim 400$~nm) and at longer wavelengths by the ring waveguide mode cutoff wavelength ($\lambda_0 \approx 2470$~nm). Our mathematical approach is illustrated in Fig.~3(a), and the resulting full spectral response of grating-induced loss is shown by the red trace in Fig.~3(b). This spectrum is obtained by transforming the variable $N/m$ into $\lambda/\Lambda$ using the relationships $\Lambda = 2 \pi (RR-RW)/N$ and $\lambda = 2 \pi (RR-RW)/{m}$, and subsequently mapping it to $\lambda_{0}$ using $ N/m =\lambda/\Lambda =\lambda_{0}/(n_\text{eff}\Lambda)$. The wavelength dependence of $n_\text{eff}$ for the TE$_0$ mode is shown in the inset plot of Fig. 3(b). These results reveal that the identified loss channels can significantly influence the performance and efficiency of wideband nonlinear photonic processes, where maintaining high $Q$ is essential. We illustrate the potential implication of grating-induced radiation loss in Figs. 3(c)–(d) by considering two nonlinear processes: OPO and four-wave mixing Bragg scattering (FWM-BS), both spanning the blue and infrared spectral regions. Specifically, we examine 493~nm and 2236~nm generation from an 807~nm pump for OPO, and frequency downconversion from 493~nm to 1545~nm for FWM-BS. We assume a PhCR design in which the mode splitting is targeted at 2236~nm (corresponding to $N/m = 2$) to phase-match the nonlinear process~\cite{Stone_NatPhoton_2024}. However, the 493~nm wavelength, i.e., of particular relevance to laser cooling transitions in barium ions, coincides with a pronounced radiation loss peak at $N/m \approx 0.33$. This spectral overlap would significantly increase the OPO threshold power~\cite{lu_milliwatt_2019} or reduce the FWM-BS conversion efficiency~\cite{li_efficient_single_photon_2016} due to this region's low-quality factor $Q$. While this loss channel could be avoided through design adjustments, the broader band of excess loss extending up to 2000~nm ($N/m \approx 1.7$) presents a general challenge for frequency mixing schemes involving long-wavelength mode splitting coupled to much shorter wavelengths. An alternative strategy is to apply mode splitting to the highest-frequency (shortest-wavelength) mode, as no excess loss is observed for $N/m > 2$. However, this approach requires the fabrication of much smaller grating periods, which introduces additional complexity, potential imperfections, and increased variability. Therefore, grating-induced radiation loss channels must be carefully considered and mitigated in the design of nonlinear photonic devices.

On the other hand, grating-induced loss can also serve as a valuable mechanism for suppressing competing nonlinear processes. In applications such as OPO and FWM-BS, it is important not only to promote the desired nonlinear processes but also to suppress undesired or parasitic processes that may consume pump power or introduce spectral noise. Recent studies have demonstrated that introducing targeted excess loss that intentionally reduces the quality factor of unwanted modes, such as through bound states-in-the-continuum~\cite{lei_hyperparametric_2023} or resonator coupling engineering~\cite{xia_energy_dissipation_2024, lu_milliwatt_2019}, can effectively suppress such competing effects. In this context, the broadband excess loss region identified in our study offers a promising design tool. By tailoring the gratings inscribed into a PhCR to align this excess loss region with the wavelengths of unwanted processes, it may be possible to passively suppress undesired nonlinear interactions across a wide spectral window. This type of loss-engineered selectivity complements PhCR-based dispersion engineering strategies and can open new routes for robust and more efficient nonlinear light generation in integrated photonic platforms.

Figure 4 presents additional simulated results showing the grating-induced loss spectrum for PhCR devices with different grating modulation amplitudes ($A = \{50,100,150\}$~nm). These results demonstrate that the loss increases with larger modulation amplitudes, consistent with the expected enhancement of mode coupling strength. Despite these variations in loss magnitude, the overall shape and $N/m$ positions of the loss channels remain largely unchanged, confirming that the phase-matching conditions and the radiation angle of OAM modes primarily govern all identified grating-induced loss channels. In addition to the previously identified peaks, we observe the emergence of small additional loss peaks ($N \approx 0.07m$ and $N \approx 1.93m$) attributed to coupling between the TE$_0$ mode and the fundamental transverse magnetic (TM$_0$) mode. This feature becomes more noticeable at higher modulation amplitudes due to increased inter-modal coupling. These findings reinforce the conclusions from earlier simulations: while grating strength impacts the overall loss magnitude, the location of critical loss channels based on $N/m$ ratio is robust to variations in grating amplitude. This further highlights the importance of carefully balancing grating strength and spectral placement when designing PhCRs for low-loss, wideband nonlinear photonic applications.

\begin{figure*}[t!]
\centering\includegraphics[width=0.75\linewidth]{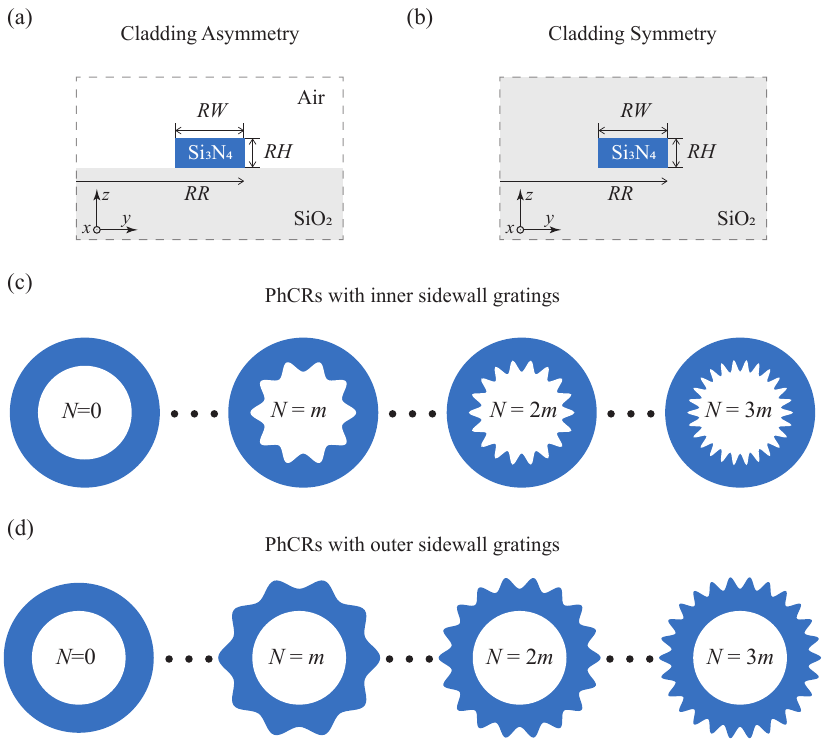}
\caption{Schematic illustrations of PhCRs with inner and outer sidewall sinusoidal gratings. (a) Cross-sectional view of a PhCR waveguide with cladding asymmetry, consisting of air cladding on top and silicon dioxide (SiO$_2$) cladding at the bottom. The waveguide core is silicon nitride (Si$_3$N$_4$), with geometric parameters consisting of its outer ring radius ($RR$), ring width ($RW$), and thickness ($RH$). (b) Cross-sectional view of a PhCR waveguide with symmetric cladding consisting of SiO$_2$ above and below the Si$_3$N$_4$ core. (c) Method for studying the full spectral response of grating-induced loss in PhCRs with inner sidewall gratings. The number of grating periods ($N$) is varied while all other geometric parameters are kept constant. (d) We used the same method to study PhCRs with outer sidewall sinusoidal gratings.
} 
\label{Figs1}
\end{figure*}

\begin{figure*}[t!]
\centering\includegraphics[width=0.75\linewidth]{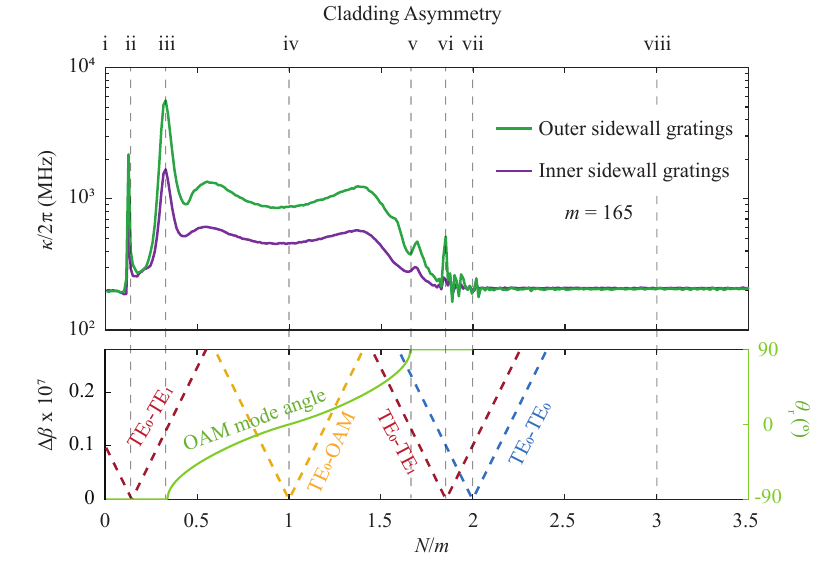}
\caption{Numerical simulation of grating-induced loss in PhCRs with outer and inner sidewall gratings under asymmetric cladding conditions. The top plot shows the full spectral response of grating-induced loss as a function of the grating-to-mode number ratio ($N/m$), comparing PhCRs with outer sidewall gratings (solid green line) and inner sidewall gratings (solid purple line). Simulations were conducted at a fixed free-space wavelength of 1550~nm, with an azimuthal mode number of $m = 165$, while $N$ varies from 0 to approximately 600. All the other geometric parameters are given as follows: an outer radius of $RR = 25~\upmu$m, a ring width of $RW = 1500$~nm, and a thickness of $RH = 500$~nm, with a grating modulation amplitude of $A = 20$~nm. The bottom plot presents the calculated radiation angle ($\theta_{\text{r}}$) for OAM modes (solid green line), along with the relative phase mismatch ($\Delta\beta$) for several coupling scenarios: TE$_0$-TE$_1$ mode coupling (red dashed line), TE$_0$-OAM mode coupling (yellow dashed line), and TE$_0$-TE$_0$ mode coupling (blue dashed line). This analysis identifies distinct loss channels, including: OAM radiation centered at region (iv), surface-mode radiation at regions (iii) and (v), TE$_0$-TE$_1$ mode forward/backward coupling at regions (ii) and (vi), minimal excess loss at regions (i) and (viii), and TE$_0$-TE$_0$ backscattering at region (vii). Dashed vertical lines mark reference points (i) through (viii), corresponding to $N \approx 0$, $0.13m$, $0.33m$, $m$, $1.67m$, $1.87m$, $2m$, and $3m$, respectively.
} 
\label{Figs2}
\end{figure*}

\begin{figure*}[t!]
\centering\includegraphics[width=0.75\linewidth]{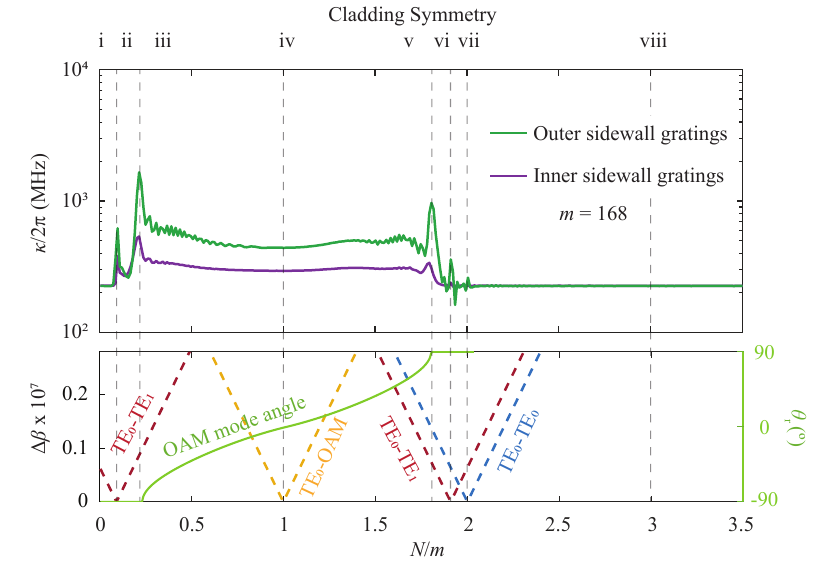}
\caption{Numerical simulation of grating-induced loss in PhCRs with outer and inner sidewall gratings under symmetric cladding conditions. The top plot shows the full spectral response of grating-induced loss as a function of the grating-to-mode number ratio ($N/m$), comparing PhCRs with outer sidewall gratings (solid green line) and inner sidewall gratings (solid purple line). Simulations were conducted at a fixed free-space wavelength of 1550~nm, with an azimuthal mode number of $m = 168$, while $N$ varies from 0 to approximately 600. All the other geometric parameters are the same as in Fig.~\ref{Figs2}. The bottom plot presents the calculated radiation angle ($\theta_{\text{r}}$) for OAM modes (solid green line), along with the relative phase mismatch ($\Delta\beta$) for several coupling scenarios: TE$_0$-TE$_1$ mode coupling (red dashed line), TE$_0$–OAM mode coupling (yellow dashed line), and TE$_0$-TE$_0$ mode coupling (blue dashed line). This analysis identifies distinct loss channels, including: OAM radiation centered at region (iv), surface-mode radiation at regions (iii) and (v), TE$_0$-TE$_1$ mode forward/backward coupling at regions (ii) and (vi), minimal excess loss at regions (i) and (viii), TE$_0$-TE$_0$ backscattering at region (vii). Dashed vertical lines mark reference points (i) through (viii), corresponding to $N \approx 0$, $0.10m$, $0.22m$, $m$, $1.78m$, $1.90m$, $2m$, and $3m$, respectively.
} 
\label{Figs3}
\end{figure*}

\section{Conclusion}
In conclusion, we have experimentally and numerically investigated the broadband spectral behavior of grating-induced loss in photonic crystal microring resonators, revealing critical insights into their impact on nonlinear photonic processes. By systematically varying the number of inscribed grating periods while keeping all other device parameters fixed, we characterized the loss spectrum as a function of the grating-to-mode number ratio ($N/m$). This approach, which eliminates the need for multiple laser sources, enabled us to capture subtle but important loss mechanisms introduced by the incorporation of gratings into microrings. Our experimental results, validated by 3D FDTD simulations, identified distinct loss channels within a broad excess-loss region associated with vertical emission into OAM modes. We analyzed the origin of these loss channels using the radiation angle of OAM modes and phase mismatch calculations. By transforming the $N/m$-dependent loss into a function of wavelength, we established the full spectral response, highlighting how grating-induced loss evolves across a broadband wavelength range in a fixed-geometry PhCR. These findings are directly relevant to a wide array of nonlinear photonic applications, including optical parametric oscillation, Kerr frequency comb generation, four-wave mixing Bragg scattering, and self-injection locking. In particular, we showed that loss channels near $N/m \approx 0.33$ can spectrally overlap with signal wavelengths in frequency-mixing schemes, reducing efficiency and increasing operational thresholds. Moreover, the broad region of loss extending up to $N/m\approx1.7$ 
presents additional challenges for desired nonlinear frequency mixing. We also discussed how this broad loss region and the specific loss channels identified in this work can be strategically leveraged as a design tool for suppressing competing nonlinear processes by aligning them with unwanted spectral components.
%Such broadband loss effects are often overlooked in PhCR device design but can critically impact performance when interacting modes span widely separated wavelengths. 
Overall, this work provides a practical framework for predicting and mitigating grating-induced losses across the full spectral bandwidth of PhCRs. Our methodology offers clear design guidelines to avoid detrimental radiation channels and to maximize nonlinear conversion efficiency, thereby paving the way for more robust and efficient integrated photonic systems operating across broad spectral ranges. %\vspace{0.2 cm}

%\medskip
\appendix
\section*{Appendix: Numerical Simulations of Inner and Outer Sidewall Gratings under Symmetric and Asymmetric Claddings}
In this section, we further investigate the full spectral response of grating-induced loss in PhCRs featuring sinusoidal gratings inscribed on either the inner~\cite{Lu_PhotonRes_2020,Black_Optica_2022, Ulanov_NatPhoton_2024} or outer~\cite{lu_selective_2014} sidewall, as illustrated in Fig.~\ref{Figs1}. For both configurations, we consider two sets of simulations: one with asymmetric cladding, consisting of air above and silicon dioxide (SiO$_2$) below the Si$_3$N$_4$ waveguide core, and the other with symmetric cladding, where SiO$_2$ surrounds the core from both the top and bottom.
As shown in Fig.~\ref{Figs2}, under asymmetric cladding conditions, PhCRs with inner and outer sidewall gratings exhibit similar spectral characteristics in terms of the locations of loss channels along the grating-to-mode number ratio ($N/m$). However, we observe that PhCRs with outer sidewall gratings exhibit significantly higher overall loss—estimated to be \ks{$\approx89~\%$} greater than that of inner gratings, within the broad excess loss region associated with orbital angular momentum (OAM) modes. This difference is likely due to the field distribution of whispering-gallery modes, which concentrate more toward the outer sidewall of the microring waveguide~\cite{Lu_NatPhoton_2022}.
Additionally, we observe suppression of loss channels (v) and (vi) relative to (ii) and (iii) in the asymmetric cladding case, likely caused by the vertical asymmetry of the surrounding refractive index environment. To explore this further, we performed the same simulations under symmetric cladding conditions (Fig.~\ref{Figs3}). In this configuration, the amplitudes of channels (ii)/(v) and (iii)/(vi) are more balanced, and the overall trend remains consistent: outer gratings still yield higher loss, approximately 50~\% more than inner gratings, within the OAM-related excess loss region, while the locations of loss peaks remain unchanged for both grating configurations.
Finally, when comparing symmetric and asymmetric cladding conditions, we find that asymmetric cladding increases the total loss by 56~\% for inner gratings and 97~\% for outer gratings, relative to the symmetric configuration. These results underscore the importance of both cladding symmetry and grating incorporation in determining grating-induced loss characteristics, and offer valuable design insights for minimizing losses in broadband nonlinear photonic devices.

%\clearpage
%\noindent \textbf{Extended data}
\renewcommand{\figurename}{Extended Data Fig.}
\renewcommand{\tablename}{Extended Data Tab.}
\setcounter{figure}{0}

\medskip
\noindent\textbf{Funding} This work is supported by the NIST-on-a-chip program.

\medskip
\noindent \textbf{Acknowledgements --} The authors acknowledge Jin Liu, Mikkel Heuck, and Jordan Stone for helpful discussions.

\medskip
\noindent \textbf{Disclosures} The authors declare no conflicts of interest.

\medskip
\noindent \textbf{Data Availability} Data underlying the results presented in this paper may be obtained from the authors upon reasonable request.

\clearpage
\bibliographystyle{osajnl}
\bibliography{PhCR_arxiv}

\begin{thebibliography}{10}
\newcommand{\enquote}[1]{``#1''}
\expandafter\ifx\csname url\endcsname\relax
  \def\url#1{\texttt{#1}}\fi
\expandafter\ifx\csname urlprefix\endcsname\relax\def\urlprefix{URL }\fi
\providecommand{\eprint}[2][]{\url{#2}}

\bibitem{Lee_OL_2012}
J.~Y. Lee and P.~M. Fauchet, \enquote{Slow-light dispersion in periodically patterned silicon microring resonators,} Opt. Lett. \textbf{37}(1), 58--60 (2012).

\bibitem{zhang_resonance-splitting_2008}
Z.~Zhang, M.~Dainese, L.~Wosinski, and M.~Qiu, \enquote{Resonance-splitting and enhanced notch depth in SOI ring resonators with mutual mode coupling,} Opt. Express \textbf{16}(7), 4621--4630 (2008). \urlprefix\url{https://opg.optica.org/oe/abstract.cfm?URI=oe-16-7-4621}.

\bibitem{lu_selective_2014}
X.~Lu, S.~Rogers, W.~C. Jiang, and Q.~Lin, \enquote{Selective engineering of cavity resonance for frequency matching in optical parametric processes,} Applied Physics Letters \textbf{105}(15) (2014).

\bibitem{Lu_Nanophotonics_2023}
X.~Lu, F.~Zhou, Y.~Sun, A.~Chanana, M.~Wang, A.~McClung, V.~A. Aksyuk, M.~Davanco, and K.~Srinivasan, \enquote{Rod and slit photonic crystal microrings for on-chip cavity quantum electrodynamics,} Nanophotonics \textbf{12}, 521--529 (2023).

\bibitem{li_using_2019}
A.~Li and W.~Bogaerts, \enquote{Using Backscattering and Backcoupling in Silicon Ring Resonators as a New Degree of Design Freedom,} Laser \& Photonics Reviews \textbf{13} (2019). \urlprefix\url{https://api.semanticscholar.org/CorpusID:165055443}.

\bibitem{wu_side_mode_suppressed_2019}
N.~Wu and L.~Xia, \enquote{Side-mode suppressed filter based on anangular grating-subwavelength grating microring resonator with high flexibility in wavelength design,} Applied Optics \textbf{58}(26), 7174--7180 (2019).

\bibitem{leonardis_performance_2014}
F.~De~Leonardis, C.~E. Campanella, B.~Troia, A.~G. Perri, and V.~M. Passaro, \enquote{Performance of SOI Bragg grating ring resonator for nonlinear sensing applications,} Sensors \textbf{14}(9), 16,017--16,034 (2014).

\bibitem{lo_phc_biosensing_2017}
S.~M. Lo, S.~Hu, G.~Gaur, Y.~Kostoulas, S.~M. Weiss, and P.~M. Fauchet, \enquote{Photonic crystal microring resonator for label-free biosensing,} Optics express \textbf{25}(6), 7046--7054 (2017).

\bibitem{ma_integrated_label_free_2016}
T.~Ma, L.~Sun, J.~Yuan, X.~Sang, B.~Yan, K.~Wang, and C.~Yu, \enquote{Integrated label-free optical biochemical sensor with a large measurement range based on an angular grating-microring resonator,} Applied Optics \textbf{55}(18), 4784--4790 (2016).

\bibitem{arbabi_realization_2011}
A.~Arbabi, Y.~M. Kang, C.-Y. Lu, E.~Chow, and L.~L. Goddard, \enquote{Realization of a narrowband single wavelength microring mirror,} Applied Physics Letters \textbf{99}(9), 091,105 (2011). \eprint{https://pubs.aip.org/aip/apl/article-pdf/doi/10.1063/1.3633111/13813748/091105\_1\_online.pdf}, \urlprefix\url{https://doi.org/10.1063/1.3633111}.

\bibitem{arbabi_grating_single_mode_laser_2015}
A.~Arbabi, S.~M. Kamali, E.~Arbabi, B.~G. Griffin, and L.~L. Goddard, \enquote{Grating integrated single mode microring laser,} Optics Express \textbf{23}(4), 5335--5347 (2015).

\bibitem{Black_Optica_2022}
J.~A. Black, G.~Brodnik, H.~Liu, S.-P. Yu, D.~R. Carlson, J.~Zang, T.~C. Briles, and S.~B. Papp, \enquote{Optical-parametric oscillation in photonic-crystal ring resonators,} Optica \textbf{9}(10), 1183--1189 (2022). \urlprefix\url{https://opg.optica.org/optica/abstract.cfm?URI=optica-9-10-1183}.

\bibitem{lu_optics_letters_2022}
X.~Lu, A.~Chanana, F.~Zhou, M.~Davanco, and K.~Srinivasan, \enquote{Kerr optical parametric oscillation in a photonic crystal microring for accessing the infrared,} Opt. Lett. \textbf{47}(13), 3331--3334 (2022).

\bibitem{Stone_NatPhoton_2024}
J.~R. Stone, X.~Lu, G.~Moille, D.~Westly, T.~Rahman, and K.~Srinivasan, \enquote{Wavelength-accurate nonlinear conversion through wavenumber selectivity in photonic crystal resonators,} Nat. Photonics \textbf{18}, 192--199 (2024).

\bibitem{li_broadband_and_accurate_2025}
J.~Li, Y.~Zhang, J.~Zeng, and S.~Yu, \enquote{Broadband and accurate electric tuning of on-chip efficient nonlinear parametric conversion,} Optica \textbf{12}(3), 424--432 (2025).

\bibitem{Lucas_NatPhoton_2023}
E.~Lucas, S.-P. Yu, T.~C. Briles, D.~R. Carlson, and S.~B. Papp, \enquote{Tailoring microcombs with inverse-designed, meta-dispersion microresonators,} Nat. Photon. \textbf{17}, 943--950 (2023).

\bibitem{liu_implementing_2025}
H.~Liu, I.~Dickson, A.~Antohe, L.~G. Carpenter, J.~Zang, A.~R. Carollo, A.~Dan, J.~A. Black, and S.~B. Papp, \enquote{Implementing photonic-crystal resonator frequency combs in a photonic foundry,} Opt. Lett. \textbf{50}(8), 2570--2573 (2025).

\bibitem{liu_integrated_vortex_2024}
Y.~Liu, C.~Lao, M.~Wang, Y.~Cheng, Y.~Wang, S.~Fu, C.~Gao, J.~Wang, B.-B. Li, Q.~Gong, \emph{et~al.}, \enquote{Integrated vortex soliton microcombs,} Nature Photonics \textbf{18}(6), 632--637 (2024).

\bibitem{chen_integrated_vortex_2024}
B.~Chen, Y.~Zhou, Y.~Liu, C.~Ye, Q.~Cao, P.~Huang, C.~Kim, Y.~Zheng, L.~K. Oxenl{\o}we, K.~Yvind, \emph{et~al.}, \enquote{Integrated optical vortex microcomb,} Nature Photonics \textbf{18}(6), 625--631 (2024).

\bibitem{Yu_NatPhoton_2021}
S.-P. Yu, D.~C. Cole, H.~Jung, G.~T. Moille, K.~Srinivasan, and S.~B. Papp, \enquote{Spontaneous pulse formation in edgeless photonic crystal resonators,} Nat. Photonics \textbf{15}(6), 461--467 (2021).

\bibitem{yu_continuum_bright_pulse_2022}
S.-P. Yu, E.~Lucas, J.~Zang, and S.~B. Papp, \enquote{A continuum of bright and dark-pulse states in a photonic-crystal resonator,} Nature Communications \textbf{13}(1), 3134 (2022).

\bibitem{Ulanov_NatPhoton_2024}
A.~E. Ulanov, T.~Wildi, N.~G. Pavlov, J.~D. Jost, M.~Karpov, and T.~Herr, \enquote{Synthetic reflection self-injection-locked microcombs,} Nature Photonics \textbf{18}(3), 294--299 (2024).

\bibitem{lu_band_flipping_2024}
X.~Lu, A.~Chanana, Y.~Sun, A.~McClung, M.~Davanco, and K.~Srinivasan, \enquote{Band flipping and bandgap closing in a photonic crystal ring and its applications,} Optics Express \textbf{32}(11), 20,360--20,369 (2024).

\bibitem{ulanov_quadrature_squeezing_2025}
A.~E. Ulanov, B.~Ruhnke, T.~Wildi, and T.~Herr, \enquote{Quadrature squeezing in a nanophotonic microresonator,} arXiv preprint arXiv:2502.17337  (2025).

\bibitem{Wang_PRX_2024}
M.~Wang, Z.-G. Hu, C.~Lao, Y.~Wang, X.~Jin, X.~Zhou, Y.~Lei, Z.~Wang, W.~Liu, Q.-F. Yang, and B.-B. Li, \enquote{Taming Brillouin Optomechanics Using Supermode Microresonators,} Phys. Rev. X \textbf{14}, 011,056 (2024).

\bibitem{cai_integrated_oam_2012}
X.~Cai, J.~Wang, M.~J. Strain, B.~Johnson-Morris, J.~Zhu, M.~Sorel, J.~L. O’Brien, M.~G. Thompson, and S.~Yu, \enquote{Integrated {Compact} {Optical} {Vortex} {Beam} {Emitters},} Science \textbf{338}(6105), 363--366 (2012).

\bibitem{willner_OAM_2021}
A.~E. Willner, K.~Pang, H.~Song, K.~Zou, and H.~Zhou, \enquote{Orbital angular momentum of light for communications,} Applied Physics Reviews \textbf{8}(4) (2021).

\bibitem{Wang_PRL_2022}
M.~Wang, F.~Zhou, X.~Lu, A.~McClung, M.~Davanco, V.~A. Aksyuk, and K.~Srinivasan, \enquote{Fractional Optical Angular Momentum and Multi-Defect-Mediated Mode Renormalization and Orientation Control in Photonic Crystal Microring Resonators,} Phys. Rev. Lett. \textbf{129}, 186,101 (2022).

\bibitem{Lu_NatCommun_2023}
X.~Lu, M.~Wang, F.~Zhou, M.~Heuck, W.~Zhu, V.~A. Aksyuk, D.~R. Englund, and K.~Srinivasan, \enquote{Highly-twisted states of light from a high quality factor photonic crystal ring,} Nat. Commun. \textbf{14}(1), 1119 (2023).

\bibitem{Lu_PhotonRes_2020}
X.~Lu, A.~Rao, G.~Moille, D.~A. Westly, and K.~Srinivasan, \enquote{Universal frequency engineering tool for microcavity nonlinear optics: multiple selective mode splitting of whispering-gallery resonances,} Photon. Res. \textbf{8}(11), 1676--1686 (2020).

\bibitem{Lu_NatPhoton_2022}
X.~Lu, A.~McClung, and K.~Srinivasan, \enquote{High-{Q} slow light and its localization in a photonic crystal microring,} Nat. Photonics \textbf{16}, 66--71 (2022).

\bibitem{Lu_PhotonRes_2023}
X.~Lu, Y.~Sun, A.~Chanana, U.~A. Javid, M.~Davanco, and K.~Srinivasan, \enquote{Multi-mode microcavity frequency engineering through a shifted grating in a photonic crystal ring,} Photon. Res. \textbf{11}(11), A72 (2023).

\bibitem{Moille_CommunPhys_2023}
G.~Moille, X.~Lu, J.~Stone, D.~Westly, and K.~Srinivasan, \enquote{Fourier synthesis dispersion engineering of photonic crystal microrings for broadband frequency combs,} Commun. Phys. \textbf{6}(1), 144 (2023).

\bibitem{luke_si3n4_2015}
K.~Luke, Y.~Okawachi, M.~R. Lamont, A.~L. Gaeta, and M.~Lipson, \enquote{Broadband mid-infrared frequency comb generation in a Si3N4 microresonator,} Opt. Lett. \textbf{40}(21), 4823--4826 (2015).

\bibitem{nitkowski_cavity_enhanced_2008}
A.~Nitkowski, L.~Chen, and M.~Lipson, \enquote{Cavity-enhanced on-chip absorption spectroscopy using microring resonators,} Optics express \textbf{16}(16), 11,930--11,936 (2008).

\bibitem{sorayaie_focus_on_surface_scattering_2025}
P.~Sorayaie, L.~Hajshahvaladi, M.~Kolahdouz, K.~Golshan, and G.-M. Parsanasab, \enquote{Design optimization for manufacturing polymer microring lasers: Focus on surface scattering losses,} Optics \& Laser Technology \textbf{182}, 112,101 (2025).

\bibitem{lee_influence_of_surface_2019}
H.~Lee, T.~Kananen, A.~Soman, and T.~Gu, \enquote{Influence of surface roughness on microring-based phase shifters,} IEEE Photonics Technology Letters \textbf{31}(11), 813--816 (2019).

\bibitem{moille_broadband_CMT_2019}
G.~Moille, Q.~Li, T.~C. Briles, S.-P. Yu, T.~Drake, X.~Lu, A.~Rao, D.~Westly, S.~B. Papp, and K.~Srinivasan, \enquote{Broadband resonator-waveguide coupling for efficient extraction of octave-spanning microcombs,} Opt. Lett. \textbf{44}(19), 4737--4740 (2019).

\bibitem{cui_distinguishing_2024}
C.~Cui, L.~Zhang, B.-H. Wu, S.~Liu, P.-K. Chen, and L.~Fan, \enquote{Distinguishing under-and over-coupled resonances without prior knowledge,} Optica \textbf{11}(2), 176--177 (2024).

\bibitem{ji_methods_2021}
X.~Ji, S.~Roberts, M.~Corato-Zanarella, and M.~Lipson, \enquote{Methods to achieve ultra-high quality factor silicon nitride resonators,} APL Photonics \textbf{6}(7), 071,101 (2021). \urlprefix\url{https://aip.scitation.org/doi/10.1063/5.0057881}.

\bibitem{Yariv}
A.~Yariv and P.~Yeh, \emph{Photonics: Optical Electronics in Modern Communications} (Oxford University Press, 2007).

\bibitem{flueckiger_subwavelength_grating_2016}
J.~Flueckiger, S.~Schmidt, V.~Donzella, A.~Sherwali, D.~M. Ratner, L.~Chrostowski, and K.~C. Cheung, \enquote{Sub-wavelength grating for enhanced ring resonator biosensor,} Optics express \textbf{24}(14), 15,672--15,686 (2016).

\bibitem{ji_compact_spatial_mode_2022}
X.~Ji, J.~Liu, J.~He, R.~N. Wang, Z.~Qiu, J.~Riemensberger, and T.~J. Kippenberg, \enquote{Compact, spatial-mode-interaction-free, ultralow-loss, nonlinear photonic integrated circuits,} Communications Physics \textbf{5}(1), 84 (2022).

\bibitem{bahadori_design_space_2018}
M.~Bahadori, M.~Nikdast, S.~Rumley, L.~Y. Dai, N.~Janosik, T.~Van~Vaerenbergh, A.~Gazman, Q.~Cheng, R.~Polster, and K.~Bergman, \enquote{Design space exploration of microring resonators in silicon photonic interconnects: impact of the ring curvature,} Journal of lightwave technology \textbf{36}(13), 2767--2782 (2018).

\bibitem{alonso_diffraction_less_2019}
C.~Alonso-Ramos, X.~Le~Roux, J.~Zhang, D.~Benedikovic, V.~Vakarin, E.~Dur{\'a}n-Valdeiglesias, D.~Oser, D.~P{\'e}rez-Galacho, F.~Mazeas, L.~Labont{\'e}, \emph{et~al.}, \enquote{Diffraction-less propagation beyond the sub-wavelength regime: a new type of nanophotonic waveguide,} Scientific Reports \textbf{9}(1), 5347 (2019).

\bibitem{lu_milliwatt_2019}
X.~Lu, G.~Moille, A.~Singh, Q.~Li, D.~A. Westly, A.~Rao, S.-P. Yu, T.~C. Briles, S.~B. Papp, and K.~Srinivasan, \enquote{Milliwatt-threshold visible--telecom optical parametric oscillation using silicon nanophotonics,} Optica \textbf{6}(12), 1535--1541 (2019).

\bibitem{li_efficient_single_photon_2016}
Q.~Li, M.~Davan{\c{c}}o, and K.~Srinivasan, \enquote{Efficient and low-noise single-photon-level frequency conversion interfaces using silicon nanophotonics,} Nature Photonics \textbf{10}(6), 406--414 (2016).

\bibitem{lei_hyperparametric_2023}
F.~Lei, Z.~Ye, K.~Twayana, Y.~Gao, M.~Girardi, {\'O}.~B. Helgason, P.~Zhao, and V.~Torres-Company, \enquote{Hyperparametric oscillation via bound states in the continuum,} Physical Review Letters \textbf{130}(9), 093,801 (2023).

\bibitem{xia_energy_dissipation_2024}
D.~Xia, J.~Zhao, H.~Cheng, Z.~Wang, J.~Huang, L.~Luo, D.~Liu, S.~Yang, B.~Zhang, and Z.~Li, \enquote{Energy dissipation engineering for widely tunable (1.2--2.1 $\mu$m) optical parametric oscillation in integrated chalcogenide microresonators,} Laser \& Photonics Reviews \textbf{18}(10), 2301,098 (2024).

\end{thebibliography}

%\appendix*
%\onecolumngrid
%\include{supp_arxiv}
\end{document}